\documentclass[fleqn]{article}
\usepackage{epsfig}
\begin{document}
\newcommand{\volume}{?}              
\newcommand{\xyear}{2001}            
\newcommand{\issue}{?}               
\newcommand{\recdate}{dd.mm.yyyy}    
\newcommand{\revdate}{dd.mm.yyyy}    
\newcommand{\revnum}{0}              
\newcommand{\accdate}{dd.mm.yyyy}    
\newcommand{\coeditor}{ue}           
\newcommand{\firstpage}{1}           
\newcommand{\lastpage}{13}            
\setcounter{page}{\firstpage}        
\newcommand{\keywords}{Gravity; gravitons; 4-dimensional optics;
  elementary particles.}
\newcommand{\PACS}{04.20.Cv, 04.60.-m, 42.25.Bs}
\markright{A theory of mass and gravity in
4-dimensional optics} 
\title{A theory of mass and gravity in
4-dimensional optics}
\author{J.\ B.\ Almeida \\ \small{Universidade do Minho, Physics Department,
 Campus de Gualtar,}\\
  \small{Braga, 4710-057,
  Portugal. \texttt{bda@fisica.uminho.pt}}}
  \date{}
\maketitle

\begin{abstract}                
The paper deals with the concepts of mass and gravity in the
formalism of 4-dimensional optics, previously introduced by  the
author. It is shown that elementary particles can be associated
with 4-dimensional standing wave patterns with the Compton
wavelength and both inertial and gravitational mass are derived
from this concept and shown to be attributable to an waveguide
laid along the particle's worldline; the same formalism is shown
to accommodate also mass due to binding energy within compact
bodies. Momentum exchange with accelerated bodies through
gravitons is discussed and shown similar to mode exchange in
optical fibers. Reported anomalies on the behaviour of the
Foucault pendulum, both periodic and exceptional on the occasion
of solar eclipses, are explained not only qualitatively but also
on order of magnitude, resorting to graviton exchanges between
Earth and the Sun or the Moon. It is argued that these effects
provide experimental evidence of gravitons.
\end{abstract}

\section{Introduction}
The paper's purpose is to show that mass and gravity can be
derived from geometrical properties of space alone. Two points of
departure are used in order to reach the same end, those being
mass scaling of coordinates introduced in a previous paper
\cite{Almeida01} and Compton wavelength for elementary particles.

In the work mentioned above, corrected and complemented in a
later paper \cite{Almeida01:4}, mass scaling of coordinates was
introduced to allow massive bodies to follow metric geodesics of
equation
\begin{equation}
    \label{eq:lagrangean}
    2 L = g_{\alpha \beta}\dot{x}^\alpha \dot{x}^\beta =1,
\end{equation}
where $L$ is the movement Lagrangian and ''dot'' indicates time
derivative\footnote{Greek letters are used for indices taking
values between $0$ and $3$ and roman letters for indices with
values between $1$ and $3$. Use is also made of indices that refer
to a specific coordinate, like $r$, $\theta$ and $\varphi$ with
spherical coordinates.}. In this formulation $g_{\alpha \beta}$
incorporates a factor $m^2$, equal to the square of the moving
body's mass, which scales local coordinates. An alternative
method would be to scale arc length as time divided by the moving
body's mass, but this approach breaks the nice symmetry of the
Universal variational principle given by $\mathrm{d}t^2 =
g_{\alpha \beta} \mathrm{d}x^\alpha \mathrm{d}x^\beta$ and was
rejected in favor of coordinate scaling.

The same papers linked Compton wavelength of elementary particles
to waves of angular frequency $\omega = m c^2 /\hbar$ associated
to those particles and propagating along their worldline. In the
previous expression $c$ is the speed of light in vacuum and
$\hbar$ is Planck's constant divided by $2 \pi$. Using
non-dimensional units mass becomes exactly equal to the frequency
associated with the particle, so it is expected that mass and
frequency are just two different views of the same
reality\footnote{Non-dimensional units are obtained dividing
length, time and mass by the factors $\sqrt{G \hbar /c^3}$,
$\sqrt{G \hbar/c^5}$ and $\sqrt{\hbar c/G}$, respectively; $G$ is
the gravitational constant. Electric charge is normalized by the
charge of the electron but this normalization will not be needed
in the present work.}. Further along the expression \emph{Compton
frequency} will sometimes be used to refer to the angular
frequency associated with an elementary particle; this will be
considered the same as the particle's mass expressed in different
units.

This paper develops the theory further, establishing a close
connection between matter, gravity and periodic oscillations of
space. The theory is based on the formation of 4-dimensional wave
patterns as a result of localized resonance modes in elementary
particles, something that other authors have already suggested
\cite{Haisch01}. Both inertia and gravity will be derived from
this simple concept and will assume that massive bodies act as
superposition of elementary particle vibrational modes, together
with modes due to orbital and vibrational frequencies within the
body. This paper is concerned solely with gravity, but the author
believes that other interactions, namely electrodynamics and
chromodynamics, will eventually be included in the theory under a
unified approach.
\section{\label{singpart}Equivalence between Compton frequency
and inertial mass}
Previous papers \cite{Almeida01, Almeida01:4} introduced a
4-dimensional space with signature 4 which in many circumstances
of interest can become Euclidean. These circumstances include
movement under the gravitational fields due to stationary bodies,
which are generally described by the metric $g_{\alpha \beta} =
m^2 n^2 \delta_{\alpha \beta}$, with $m$ the inertial mass of a
moving body and $n$ the space curvature due to gravity. What this
metric tells is that in order to use time interval to measure
geodesic arc length one must use a local scale factor $n$
associated with an inertia scale factor $m$. It is possible to use
unscaled coordinates if the geodesic arc length is no longer
directly associated with time but rather $\mathrm{d} s =
\mathrm{d} t/m n$. The local scale factor does not represent
intrinsic curvature, but is rather a convenience that one can
choose to use or not. There are situations, naturally, when space
cannot be flattened by a simple change of coordinates; these
include those cases when the field source is not stationary,
namely when it is accelerated. The works mentioned above discuss
the electromagnetic field due to a moving charge, as one situation
where a change of coordinates would not flatten space.

It is natural to address the simple cases first and defer the
complications to later discussions. This paper proceeds in that
line, starting with the field and associated metric due to a
stationary elementary particle and generalizing the conclusions in
successive steps. The premises are the Universal measurement of
geodesic arc length provided by time intervals, if scaled
coordinates are used, and the association of a Compton frequency
 with every elementary particle.

According to the theory developed in Ref.\ \cite{Almeida01}, the
4-dimensional worldline of a particle with mass $m$ is given by
Eq.\ ({\ref{eq:lagrangean}). If the particle is under the
influence of a stationary gravitational field the metric takes
the form $m^2 n^2 \delta_{\alpha \beta}$, with $n$ a function of
the spatial coordinates $x^i$, ($i=1,2,3$). Considerations made
in the next section justify the allowance that is made here for
the gravitational field $n$ of a stationary body to be a function
of all the 4 coordinates without changing the diagonal form of the
metric. Defining the conjugate momentum $k_\alpha =
\partial L/
\partial \dot{x}^\alpha$ it is
\begin{equation}
    \label{eq:momentum}
    k_\alpha = g_{\alpha \beta} \dot{x}^\beta = m^2 n^2
    \delta_{\alpha \beta}\dot{x}^\beta,
\end{equation}
which allows the geodesic equation to re-written in terms of
momentum components
\begin{equation}
    \label{eq:altgeodesic}
    \delta^{\alpha \beta}k_\alpha k_\beta = m^2 n^2.
\end{equation}

This equation remains unchanged if  both sides are multiplied by
the harmonic wave function $\psi = \mathrm{exp}
(\mathrm{j}\,k_\alpha x^\alpha )$. This wave function represents
a pattern of standing plane waves in 4-dimensional space but it is
a truly propagating wave in 3D; in fact  spatial dependence can
be separated from the dependence on $x^0$ as $\psi =
\mathrm{exp}[\mathrm{j}\,(k_0 x^0 + k_i x^i)]$, to highlight that
when $x^0$ is fixed the wave exhibits a sinusoidal variation
along the direction defined by $k_i$. The 4-dimensional momentum
$k_\alpha$ functions as wave vector for the stationary wave
pattern.

If both sides of Eq.\ (\ref{eq:altgeodesic}) are multiplied by
$\psi$, noting that $\partial_\alpha \psi \partial_\beta \psi =
\psi \partial_{\alpha \beta}\psi$, one gets the harmonic wave
equation
\begin{equation}
    \label{eq:wave}
    \delta^{\alpha \beta} \partial_{\alpha \beta} \psi
    = -m^2 n^2 \psi,
\end{equation}
indicating that the wave pattern has a spatial frequency $m n$
along the wavefront normal.

The equation can have a different interpretation if  a new wave
function is introduced as $\Psi = m \psi$. Then $m^2 = \Psi
\Psi^*$, with $^*$ standing for complex conjugate, and he new
equation is
\begin{equation}
    \label{eq:wave2}
    \frac{\delta^{\alpha \beta} \partial_{\alpha \beta} \Psi}{\Psi
    \Psi^* n^2} = -\Psi.
\end{equation}
It is possible to say that the spatial frequency is always unity
if coordinates are scaled by mass $\sqrt{\Psi \Psi^*}$ and field
$n$. While the first interpretation is consistent with the
definition of Compton wavelength, the latter is more in line with
the author's previous work.

It is now convenient to consider a more general situation where
 the metric is allowed to have an arbitrary form. It is always
 possible to make $g_{\alpha \beta} = m^2 n_{\alpha \beta}$
 and replace Eq.\ (\ref{eq:altgeodesic}) by
\begin{equation}
    \label{eq:altgeodesic2}
    n^{\alpha \beta} k_\alpha k_\beta = m^2,
\end{equation}
with $n^{\alpha \beta}=(n_{\alpha \beta})^{-1}$. Introducing the
wave function $\Psi$ as before,
\begin{equation}
    \label{eq:wave3}
    \frac{n^{\alpha \beta} \partial_\alpha \Psi \partial_\beta
    \Psi}{\Psi \Psi^*} = - \Psi,
\end{equation}
which is consistent with the definition of a Lagrangean density by
\begin{equation}
    \label{eq:lagdens}
    \mathcal{L} = \frac{n^{\alpha \beta}
    \partial_\alpha \Psi \partial_\beta \Psi}{\Psi \Psi^*} -
    \Psi^2.
\end{equation}
The corresponding Euler-Lagrange equation is \cite{Jose98}
\begin{equation}
    \label{eq:wave4}
    \partial_\alpha \left(\frac{n^{\alpha \beta}}{ \Psi
    \Psi^*}\,\partial_\beta \Psi \right)=
    \partial_\alpha \partial^\alpha \Psi = -\Psi.
\end{equation}
This is the general form of wave equation (\ref{eq:wave2}).

Eqs.\ (\ref{eq:altgeodesic2}) and (\ref{eq:wave4}) represent two
different views of the same phenomenon; the first one describes a
particle's worldline, while the latter describes an equivalent
wave pattern and can be seen as the analogous to Klein-Gordon
equation in 4-dimensional optics. This is not unlike ray and wave
descriptions of optics, which are equivalent as long as the
dimensions involved remain large compared to the wavelength.
Similarly in 4-dimensional optics one is allowed to deal with
particle's worldlines as long as all the dimensions involved are
large compared to their Compton wavelengths.

It is possible to conclude that an elementary particle with known
momentum can be associated with a 4-dimensional stationary wave
with spatial angular frequency equal to the particle's mass
multiplied by the local gravitational field and wave vector equal
to its momentum.
\section{\label{graviton}Vacuum, gravitons and photons}
General relativity accepts that space is curved by gravity and
moving bodies are affected by space curvature. The assumption of
4-dimensional optics is that not only the gravitational field but
also the inertial mass of a moving body determine curvature, the
latter through coordinate scaling. It has been shown before
\cite{Almeida01, Almeida01:4} that electromagnetic fields can be
assigned to space curvature, which is then determined also by the
electric charge of the moving particle. The present paper is
concerned mainly with gravity and so considerations about
electromagnetic fields will not be extended; it is important to
understand, though, that ultimately all gravitational and
electromagnetic fields result from the superposition of
electrostatic and gravitational fields due to elementary
particles. The cited works showed that Lorentz force can
effectively be deduced from the electrostatic field of a moving
charge. Accordingly the effect of a moving electrically charged
elementary particle on empty space can be examined and it can be
accepted that the latter must be filled with a superposition of
similar effects.

In Ref.\ \cite{Almeida01:4} the author established the fields due
to both gravity and electric charge of a body of mass $M$ and
charge $Q$ as $n = \exp(M/r)$ and $v = \exp(Q/r)$, respectively.
Further along in this paper  discusses how the gravitational field
is generated, while the electrostatic field will be the subject
of future work; for now it is useful to accept the expressions
above just as a result of compatibility with Newtonian and
electrostatic forces. If another body with mass $m$ and electric
charge $q$ is under the influence of those fields, its movement
follows the geodesic of the space defined by the metric
\begin{equation}
    \label{eq:genmetric}
    g_{\alpha \beta} = m^2 \left[\begin{array}{cccc}
      e^{2(m M + q Q)/m r} & 0 & 0 & 0 \\
      0 & e^{2M/r} & 0 & 0 \\
      0 & 0 & e^{2M/r} & 0 \\
      0 & 0 & 0 & e^{2M/r} \
    \end{array}\right].
\end{equation}
It is convenient to decompose the metric into four components as
$g_{\alpha \beta} = m^2 n_{\alpha \beta} (v_{\alpha
\beta})^{q/m}$, where $n_{\alpha \beta}$ designates the
gravitational field, $v_{\alpha \beta}$ the electric field, $m$
is the inertial mass and $q$ the electric charge.

There are some important consequences of the equation above. First
of all  notice that the metric due to electrostatic and
gravitational fields is diagonal and can have an anisotropy on the
0th element if the electrostatic field is present. A stationary
body could be the source of an electromagnetic field but this is
never the case with an elementary particle. Notice also that the
electric charge of the body that suffers the influence of the
electrostatic field is equally responsible for the anisotropy,
while its mass influences the whole metric. In fact the metric
ceases to exist if there is no inertial mass. This is the result
of the concept of metric linked to the movement and not to space
itself. A further point that needs to be raised is that the
fields don't die away as distance increases, but rather tend
exponentially to unity, leading to a concept of a field filled
vacuum, entirely compatible with the uncertainty principle and
postulates by other authors \cite{Haisch01, Nernst16}.

The fact that inertial mass is essential for the existence of a
movement metric is shown in the wave equation (\ref{eq:wave4}) by
its collapse when $\Psi$ has zero amplitude. The question tha
must be addressed is the possibility of existence of some type of
wave solutions in vacuum which don't require mass. One can have a
particle approach similar to what was done for photons in Refs.\
\cite{Almeida01, Almeida01:4} or a wave approach which is done
below.

Consider a body following its worldline where it is possible
evaluate the derivatives $\breve{x}^i = \mathrm{d}x^i /\mathrm{d}
x^0$. When this body is stationary it is the source of a field
which is here restricted to gravity and designated $n_{\alpha
\beta}= n^2 \delta_{\alpha \beta}$; the evaluation of the field
when the body is moving involves the consideration of the tensor
\begin{equation}
    \label{eq:transform}
    {\Lambda^{\bar{\mu}}}_\nu = \left[\begin{array}{cccc}
      1 & 0 & 0 & 0 \\
      -\breve{x}^1 & 1 & 0 & 0 \\
      -\breve{x}^2 & 0 & 1 & 0 \\
      -\breve{x}^3 & 0 & 0 & 1 \
    \end{array} \right],
\end{equation}
where the "bar" over an index indicates coordinates of the moving
frame. The moving frame is taken  to be the moving body's frame
and the field on this frame is designated by $n_{\bar{\mu}
\bar{\nu}}$; the field on the stationary frame is given by
\begin{equation}
    \label{eq:transform2}
    n_{\alpha \beta} = {\Lambda_\alpha}^{\bar{\mu}}
    {\Lambda_\beta}^{\bar{\nu}} n_{\bar{\mu} \bar{\nu}};
\end{equation}
making the substitutions one gets
\begin{equation}
    \label{eq:movmetr2}
n_{\alpha \beta} = n^2 \left[\begin{array}{cccc}
      1 & -\breve{x}^1 & -\breve{x}^2 & -\breve{x}^3 \\
      -\breve{x}^1 & 1+(\breve{x}^1)^2
      & \breve{x}^1\breve{x}^2 & \breve{x}^1\breve{x}^3 \\
      -\breve{x}^2 & \breve{x}^1\breve{x}^2
       & 1+(\breve{x}^2)^2
       & \breve{x}^2\breve{x}^3 \\
      -\breve{x}^3 & \breve{x}^1\breve{x}^3
       & \breve{x}^2\breve{x}^3 & 1
       +(\breve{x}^3)^2 \
    \end{array}\right].
\end{equation}

Evaluating $n^{\alpha \beta}$:
\begin{equation}
    \label{eq:invfield}
    n^{\alpha \beta} = \frac{1}{n^2} \left[\begin{array}{cccc}
      1 + \delta_{i j}\breve{x}^i \breve{x}^j &  \breve{x}^1
       &  \breve{x}^2 &  \breve{x}^3 \\
       \breve{x}^1 & 1 & 0 & 0 \\
       \breve{x}^2 & 0 & 1 & 0 \\
       \breve{x}^3 & 0 & 0 & 1 \
    \end{array}\right];
\end{equation}
recalling Eq.\ (\ref{eq:wave4}) and passing $\Psi \Psi^*$ to the
second member  the wave equation of a zero mass field is obtained
\begin{equation}
    \label{eq:zerowave}
    \partial_\alpha \left(n^{\alpha \beta} \partial_\beta \phi \right) =
    0.
\end{equation}

A gravitational field $n^{\alpha \beta}$ is of the form given by
Eq.\ (\ref{eq:invfield}), even if it is the result of a
superposition of many individual gravitational fields; naturally
in the latter case the $\breve{x}^i$ must be replaced by
something that results from the combined movements of all the
field sources. Eq.\ (\ref{eq:zerowave}) can be expanded as
\begin{equation}
    \label{eq:zerowave2}
    \partial_0 \left(n^{0 j} \partial_j \phi \right)
    + \partial_\alpha \left( n^{\alpha 0} \partial_0 \phi \right)
    + \frac{1}{n^2} \delta^{ij} \partial_{ij}  \phi =0.
\end{equation}

It can be shown that the equation does not hold any solutions of
interest when the field is stationary, i.e.\ when the field is
created by a body in uniform motion. For an accelerated body it
is possible to try a solution on 3-space, by setting $x^0 = 0$;
if one is interested on a 3-space solution it is possible also to
try a tangent function $\Phi$, which does not depend on $x^0$ but
has the same dependence on the spatial coordinates as $\phi$. The
resulting equation is
\begin{equation}
    \label{eq:tansol1}
    \partial_0 n^{0 j} \partial_j \Phi + \frac{1}{n^2} \delta^{ij}
    \partial_{ij} \Phi =0.
\end{equation}

A plane wave type solution requires $\Phi = \exp(\mathrm{j}\, k_i
x^i)$  and leads to
\begin{equation}
    \label{eq:tansol2}
    \mathrm{j}\, \partial_0 n^{0j} k_j - \frac{\delta^{ij}k_i
    k_j}{n^2}=0.
\end{equation}
For simplicity it is  assumed that the body follows a circular
orbit with radius $r$ and proper time angular speed $\omega$
around the $x^3$ axis. Accordingly $x^1 = \cos(\omega x^0)$, $x^2
= \sin (\omega x^0 )$, $x^3 = 0$; in dealing with waves it is
convenient to write $x^1 = \mathrm{R}[\exp(\mathrm{j}\, \omega
x^0)]$ and $x^2 = \mathrm{I}[\exp(\mathrm{j}\, \omega x^0)]$,
with $\mathrm{R}()$ and $\mathrm{I}()$ meaning real part and
imaginary part, respectively. With this notation it is
$\breve{x}^1 = \mathrm{j}\, r \omega \cos(\omega x^0)$ and
$\breve{x}^2 = \mathrm{j}\, r \omega \sin(\omega x^0)$. Making
the replacements the equation becomes
\begin{equation}
    \label{eq:tansol3}
    -r \omega \left[k_1 \cos \left(\omega x^0 \right)
    + k_2 \sin \left(\omega x^0 \right) \right] -\delta^{i j} k_i k_j=
    0,
\end{equation}
which has a solution
\begin{eqnarray}
    \label{eq:ks}
    k_1 &=& -r \omega \cos \left(\omega x^0 \right), \nonumber \\
    k_2 &=& -r \omega \sin \left(\omega x^0 \right), \\
    k_3 &=& 0\,. \nonumber
\end{eqnarray}

It is legitimate to say that an orbiting point mass is the source
of a wave of frequency $\omega$ and amplitude $r \omega$ that
propagates towards the center of the orbit. A massive body is
composed of many elementary particles and orbital movements so,
in general, it is expected that a massive body is the source of
waves, or gravitons when quantization is introduced, with a
spectral distribution that results from the masses of the
individual elementary particles. These waves propagate in all
directions of space, interfering with other bodies which are the
source of the gravitational field that determines the first
body's worldline. In the particular case of two bodies orbiting
each other in circular orbits, they generate waves of equal
frequency and opposing phase which cancel each other or, to put
it in terms of gravitons, they interchange gravitons with equal
total momentum.
\section{\label{compact}Inertial mass of orbital systems}
It is now appropriate to consider an orbiting elementary particle
and search for the mass field solution of the wave equation
(\ref{eq:wave4}) which yields the inertial mass of this particle
as part of an orbital system instead of searching for massless
solutions as was done in the previous section. Without loss of
generality one can use a frame fixed to the center of the orbital
system, so as to set the whole system stationary. In this frame
the elementary particle is described by a wave $\Psi = m
\exp(\mathrm{j}\, k_\alpha x^\alpha)$, which obeys Eq.\
(\ref{eq:wave4}) with the field given by Eq.\
(\ref{eq:invfield}). The general solution of the equation is
rather complex but there is special interest on a tangential
solution valid on the origin, so only the 0th component is
examined. It is legitimate  to do this if  the particle is part
of an orbiting system that cancels out all 3-space components. The
wave equation on the origin is then
\begin{equation}
    \label{eq:inmass1}
    \partial_\alpha \left(n^{\alpha 0} \partial_0 \Psi \right) = -m^2 \Psi.
\end{equation}
If the body is in circular motion and setting
$\breve{x}^i=\mathrm{j}\, \omega x^i$, as before, one gets
\begin{equation}
    \label{eq:inmass2}
    \left(1 - r^2 \omega^2 \right)\left(k_0 \right)^2 = n^2 m^2.
\end{equation}

The value of $k_0$ can be obtained from the particle's worldline
equation $M^2 n^2 \delta_{\alpha \beta} \dot{x}^\alpha
\dot{x}^\beta$, with $M$ the particle's mass. This equation can
be set in spherical coordinates for a circular orbit as
\begin{equation}
    \label{eq:partwrldln}
    M^2 n^2 \left[ \left(\dot{x}^0 \right)^2 + \dot{r}^2 + r^2
    \dot{\varphi}^2 \right] = 1,
\end{equation}
where $\dot{\varphi}= \omega \dot{x}^0$. Replacing and solving for
$\dot{x}^0$
\begin{equation}
    \label{eq.dotx0}
    \dot{x}^0 = \frac{1}{M n \sqrt{1 + r^2 \omega^2}};
\end{equation}
inserting into Eq.\ (\ref{eq:inmass2})
\begin{equation}
    \label{eq:inmass3}
    m = M \sqrt{ \frac{1- r^2 \omega^2}{1 + r^2 \omega^2}}\,.
\end{equation}

It is important to compare the result of Eq.\ (\ref{eq:inmass3})
with the predictions of general relativity; in order to do this
one takes the first two terms of the series expansion, whereby the
inertial mass can be seen to be approximately $M (1 - r^2
\omega^2)$. Note that $\omega$ is a derivative with respect to
$x^0$ and so the linear speed is $v = r \omega \dot{x}^0$ and
taking $\dot{x}^0 = 1/\sqrt{1/n^2 - v^2}$ from Eq.\
(\ref{eq:lagrangean}) with the necessary substitutions, it is
finally $m = M (1-v/ \sqrt{1/n^2 - v^2})$. In first
approximation, the mass is equivalent to the particle's mass
reduced by an amount equal to the sum of potential and kinetic
energies.

As conclusion to the present section one can say that the inertial
mass of a compact body is the sum of all its elementary
constituents masses minus a contribution of masses resulting from
all the orbital frequencies within the body. On the wave picture,
a complex body acts as a complex wave pattern resulting from the
superposition of many harmonic wave patterns.
\section{Gravity as an evanescent field}
In section \ref{singpart} the inertial mass of an elementary
particle was associated to its Compton frequency, through a field
$\Psi$ whose nature was unspecified and in section \ref{compact}
it was shown that a compact body's mass can be viewed as a
superposition of many harmonic waves due to each of the
elementary particles' masses and to the multitude of orbital
frequencies within the body. In section \ref{graviton} it was
shown also that an orbiting body must exchange momentum with the
metric in order to follow a circular orbit and necessarily this
conclusion could be extended to any accelerated movement. All the
previous discussions were centered on inertial mass and it has not
yet been established how a particle or a body affects the metric
in order to allow this momentum exchange.

The fact that an orbiting body's mass is reflected in the center
as a different mass, when applied to an elementary particle is
indicative that containment or localization of a particle
determines its effective mass. How an elementary particle's mass
is generated is not known but it is possible to assume that it is
the result of some containment of yet another wave of different
frequency and so one  speculates that all mass probably results
from some sort of containment of harmonic waves. Eventually one
may find that all mass is the result of containment of a single
frequency which would then deserve to be designated by Higgs
frequency. Containment can be generated by a local change of the
scale factor $n$, acting as a 4-dimensional refractive index, but
it can also result from more complex metric changes. In this
framework a body or even an elementary particle is seen as a
4-dimensional waveguide extended along the body or particle's
worldline.

In a similar way to 3-dimensional optical waveguides, namely
optical fibers, the guided field originates an external
evanescent wave with the same spatial frequency along the
waveguide direction as exists inside. The following paragraphs
set the equations for this evanescent wave in the case of an
elementary particle and it will not be difficult to extend the
conclusions to more general situations. The present analysis deals
solely with the radially symmetric component of the evanescent
field, although a guided wave is expected to produce an
evanescent field due to the circular component of the wave
vector. The latter evanescent field is probably connected with
spin and electric charge and is thus outside the scope of the
present paper.

The radially symmetric component of the evanescent wave due to a
stationary particle of mass $m$ must exhibit a frequency $m$
along the 0th direction and has an equation
\begin{equation}
    \label{eq:genwave}
    v^2 \delta^{i j} \partial_{i j} \psi = -m^2 \psi,
\end{equation}
where the letter $v$ on the equation designates a propagation
speed of wavefronts defined generally as the derivative
$\mathrm{d} s /\mathrm{d} t$, with $\mathrm{d} s$ the arc length
of the wavefront normal in flat Euclidean space. Notice that this
equation could also be applied to the mass of an orbiting
particle given by Eq.\ (\ref{eq:inmass3}).

Naturally the resulting field must have spherical symmetry, which
implies that the wave equation will have a more manageable form
in spherical coordinates. Furthermore, because  $\psi$ is a
function of $r$ and $x^0$ alone, it is possible to express $v$ as
\begin{equation}
    \label{eq:genveloc}
    v = \frac{\partial_0 \psi}{\partial_r \psi}.
\end{equation}

The operator $\delta^{i j} \partial_{i j}$ is a Laplacian;
considering spherical symmetry one can make the replacement
$\delta^{i j}\partial_{i j} =\partial_{rr} + 2
\partial_r  / r $. Re-writing Eq.\ (\ref{eq:genwave})
in spherical coordinates and inserting Eq.\ (\ref{eq:genveloc})
one gets upon simplification
\begin{equation}
    \label{eq:gravwave}
    \psi
     \left(\partial_{r r} \psi + 2 \frac{\partial_r \psi}{ r} \right)
     = \left(\partial_r \psi \right)^2,
\end{equation}
which has the general solution
\begin{equation}
    \label{eq:gravwave2}
    \psi = C_1 \mathrm{e}^{(C_2/r \pm \mathrm{j}\, m x^0)},
\end{equation}
where $\mathrm{j}=\sqrt{-1}$.

So far  no comments were about the nature of the field $\psi$ but
this question must be addressed if in order to understand its
relation to gravity. It is postulated that $\psi$ is the local
coordinate scale factor, by which it is  meant that space is
corrugated with the Compton frequency on the particle's worldline
and that this corrugation is extended to infinity on the form of
an evanescent field. There must be a transition from the field on
the worldline to the evanescent field but so far there are no
means to choose among the many possibilities. In any case a
particle will always act as a 4-dimensional waveguide for the
field $\psi$, which will allow the extrapolation of many effects
known in their 3-dimensional counterparts.

The field $\psi$ defines the local scale factor or alternatively
it defines how the geodesic arc length should be measured;
accordingly in Eq.\ (\ref{eq:momentum}) one makes the assignment
$n = \sqrt{\psi \psi^*}$. In the absence of mass it is expected
that the scale factor will be unity and so constant $C_1$ in the
equation above can be made unity; constant $C_2$ must become zero
for zero mass. The field does not die away completely but an
oscillation with unit amplitude is extended to infinity; this is
seen as one possible source of quantum vacuum fluctuations
required by the uncertainty principle or the zero point field as
proposed as early as 1916 \cite{Haisch01, Nernst16}. The actual
value for constant $C_2$ is easy to establish resorting to
compatibility with Newton mechanics. If this path is taken
constant $C_2$ can be made equal to the mass, in a similar way to
what was used in Refs.\ \cite{Almeida01:4, Almeida01:2}. This
argument will be used in the present work and an independent
derivation of this constant's value will be deferred until there
is better understanding of the waveguiding process. Consequently
the gravitational field due to a stationary elementary particle
will be written as
\begin{equation}
    \label{eq:elementgrav}
    \psi = \mathrm{e}^{(m/r \pm \mathrm{j}\,m x^0)}.
\end{equation}

It is now easy to understand the mechanism of momentum exchange
by accelerated particles discussed previously, through a parallel
with 3-dimensional waveguides. It is well known that some guided
modes in an index-difference waveguide, such as an optical fiber,
can be lost when the waveguide is bent \cite{Unger77}. The
reverse is also true, although less common; a bent waveguide can
gain energy from the outside, which becomes guided energy if
further along the waveguide is straightened up. Similarly,
elementary particles are seen as 4-dimensional waveguides where
similar processes can occur. Accelerated particles have curved
worldlines which correspond to bent waveguides and are able to
exchange momentum with other particles these being the ultimate
field sources.
\section{Gravitational shielding and Foucault's pendulum}
This section makes use of the anomalies of Foucault's pendulum
oscillation reported by Allais \cite{Allais59} as experimental
evidence of gravitons. Some references will also be made to the
later observations with a torsion pendulum \cite{Saxl71} and to
the 1990 experiment in Finland, which did not confirm the previous
results \cite{Ullakko91}.

The point of departure is that gravity is carried by massless
gravitons, which are momentum carriers such as photons. The
distinction between photons and gravitons is  a question of
function and  essence which will be discussed in future work.
Gravitons, like photons, are expected to have essentially straight
trajectories that can be slightly bent by gravitational fields.
The latter effect is too small to be detected in normal
experiments and will not be considered. Gravitons must interact
with matter if they are to interchange momentum with it. So in
the majority of cases massive bodies must be considered opaque to
gravitons, although it is conceivable that in some cases they
could be transparent.

It will be shown below that all the anomalies reported by Allais
\cite{Allais59} can be explained by shielding of Solar and Lunar
gravity by the Earth or the Moon. Allais reported two periodic
anomalies, with 24h and 25h periods respectively, and a sporadic
anomaly during the Solar eclipse of 1959. It is  intended to show
that the 24h period anomaly can be explained by Earth shielding of
Solar gravity, a similar thing happening with the 25h period
anomaly but this time due to shielding of Lunar gravity. The
Solar eclipse anomaly can be explained by Lunar shielding of
Solar gravity.

The main effect of Terrestrial gravity on the pendulum is
precisely the oscillatory motion. Azimuth change of Foucault's
pendulum is due to minor differences in the direction of
gravitational pull in neighboring points on Earth. Naturally, the
fact that the Earth is rotating has the consequence that the
pendulum oscillation plane must rotate in order to eliminate the
''drag'' caused by the passage of the differential gravity.
Foucault's effect is easier to understand if one thinks of a
referential which is not subject to Earth's rotation; this
approach, however, is not the best for the comparisons made below
and so the discussion will be set on a referential on Earth.

Allais dismissed Solar and Lunar effects as possible explanation
for the observed anomalies as follows \cite{Allais59}:

\emph{ These effects are so small that none of the 19th-century
authors who worked on the theory of the pendulum, some of whom
were excellent mathematicians, ever had a desire to compute them.}

\emph{ The extreme smallness of the effects computed can readily
be accounted for if we allow for the fact that, in order to obtain
the true gradient $\vec{f}$ of the Moon and Sun attraction at a
point on the surface of the ground, with respect to Earth, we
must take the difference between the attractions at this point
and at the center of the Earth, respectively. Gradient $\vec{f}$
is of the order of $10^{-8}$.}

\emph{ Furthermore, the plane of oscillation of the pendulum can
rotate, under the influence of the solar and lunar attraction,
only because of the variations of the gradient about the point
considered. Therefore, the difference $\Delta \vec{f}$ between the
value of $\vec{f}$ at the mean position of the pendulum and its
magnitude at a nearby point must be considered. It is of some
$10^{-13}$.}

\emph{ Furthermore, nothing in the current theory of gravitation
can be considered likely to account for the screen phenomenon
observed during 1954. }

The objective is to show that if gravity screening is allowed,
solar and lunar gravitional fields have effects which are of the
same order of magnitude as terrestrial ones. In the following
S.I. units are used instead of non-dimensional ones, so that the
values obtained in the calculations are easy to compare with
everyday results.

Using spherical coordinates in a frame fixed to Earth with origin
at its center, the gravitational field can be represented by the
vector $\vec{g} = g \hat{r}$, where $g$ is the acceleration of
gravity $g = 9.8~\mathrm{m s}^{-2}$ and $\hat{r}$ is the unit
vector along the radial direction. If two points on Earth are
separated by an angle $\mathrm{d}\phi$, there is a gravity
difference between those two points whose magnitude is given by
$\mathrm{d}g = g \mathrm{d}\phi$.

The value of Sun's gravity on Earth is easily evaluated using
Newton's formula $G m/d^2$ and it averages $5.9 \times
10^{-3}~\mathrm{m s}^{-2} $. Considering shielding, that value
corresponds to the gravity on the illuminated portion of the
planet, while on the dark side it must be zero. The transition
zone from full to zero solar gravity is remarkably small and is
responsible for an appreciable gradient, in spite of the
comparatively small value of solar gravity. Dividing Sun's
diameter by the average distance from Sun to Earth one gets an
angle of $9.3 \times 10^{-3}$ radians, which corresponds roughly
to $55 \mathrm{Km}$ on Earth's surface. Dividing the value of
solar gravity by the transition zone angle yields an angular
dependence of $0.63~\mathrm{m s}^{-2}$ per radian, which is
approximately $1/15$ of terrestrial gravity variation and
perfectly in line with Allais' findings. This can easily account
for the 24h period of the anomalies.

Similar calculations performed for the Moon lead to a gravity
variation of $0.0037~\mathrm{m s}^{-2}$ per radian, which is
about $250$ times smaller than terrestrial variation. Although
the value obtained is about one order of magnitude smaller than
would be necessary to account for the observed effect, the period
is 25h as desired.

The eclipse anomaly was rightly attributed by Allais to a
screening effect. In this case the angular dependence is
considerably smaller than in the daily variation, because the
transition zone is much wider than the mere $55~ \mathrm{Km}$ of
the latter. The felt effect is of the same order of magnitude or
even larger because the linear speed of the transition zone is
very high.

Consider now the torsion pendulum experiments \cite{Saxl71,
Ullakko91}, one of which apparently confirmed Allais' results and
the other could not detect any anomalies. The second experiment
was conducted by Ullakko et al. in Helsinky during the 1970 total
solar eclipse. Naturally, both the size of the transition zone
and its speed are latitude dependent. At high latitudes the
transition zone is wider and its speed is smaller, so it is
normal that the manifestation of solar gravity variation is more
difficult to detect. This is true also for the 24h and 25h period
anomalies as well.

The effects of solar and lunar gravity variations on Foucault's
pendulum are easier to explain than on a torsion pendulum. In
fact, the period of a torsion pendulum is determined by the
suspended mass and the spring constant of the suspension. The
effect of a gravitational change in the period can be due both to
the variation in the weight of the suspended mass and on some
change of the spring constant due to extension or contraction.
These effects are necessarily much smaller than azimuthal changes
in Foucault's pendulum, where a change in period is also expected
but is probably too small to be detected. Saxl reports a period
increase during the eclipse, which was not recovered after the
eclipse was over. This  can only b attributed to creep on the
suspension caused by the gravitational change during the eclipse.

The question arises about what sort of experiments could be
conducted on Earth to prove or disprove gravity shielding theory.
Why can't one just use any sort of screen to shield form Earth's
gravity, for instance? The reason is that any screen at a fixed
height from the surface will absorb and shed an equal amount of
momentum and so gravity shielding does not have any effect. The
screen actually seems transparent to gravity. Things will be
different behind a free falling object or any object on a
geodesic orbit around Earth, for in this case the object will
absorb all the momentum that reaches it and will incorporate this
momentum in its own. Objects in free fall or in geodesic orbits
are expected to cast a gravitational shadow behind them and
effectively shield any other objects in that shadow from Earth's
gravity.
\section{Conclusions and further developments}
4D optical theory was used to successfully explain inertial and
gravitational mass. In this process it was shown that the
worldline equation for an elementary particle could be
transformed into a wave equation with the Compton spatial
frequency. It was also shown that the gravitational field could
be derived from an evanescent wave equation if it was assumed that
the particle acted like a 4-dimensional waveguide. Particles or
bodies constrained to a region of space, namely to an orbit, were
found to reflect their own mass on the center of gravity, as well
as a negative mass resulting from the binding energy.

The author was able to show that massless particles (gravitons and
photons) have wave equations similar to elementary particles but
while the latter are standing wave patterns, those of massless
particles are propagating waves in 3-dimensional space supported
on an oscillatory 0th coordinate of 4-space. Massless particles
act as momentum carriers for accelerated particles and the
momentum exchange process was shown to be similar to mode loss
and gain in optical fibers.

Allais' experimental results with a Foucault pendulum
\cite{Allais59} were used as evidence of graviton existence and
the order of magnitude of most of his reported anomalies was
explained through terrestrial and lunar shielding of gravitons
origination on the Sun and the Moon.

The author expects to have shown sufficient evidence that
elementary particles are indeed 4-dimensional waveguides, because
this hypothesis proves entirely satisfactory for the explanation
of their mass, be it inertial or gravitational mass. It is not
clear what sort of wave is being guided nor what is the guidance
mechanism. No comments were made about electromagnetism or weak
and strong interactions, although the author has some hope that
the future will allow the derivation of all the elementary
particles as guided modes of one single frequency wave (the Higgs
frequency) and all the interactions as evanescent fields relative
to these modes. Some preliminary results have already shown that,
at least qualitatively, this might be true.
\section{Acknowledgements}
The author wishes to acknowledge the many improvements to the
paper that resulted from discussions with his brother Luis B.
Almeida from INESC, Lisbon. His criticism and insistence were
decisive in the correction of several mistakes and in making the
paper understandable.

  \bibliography{aberrations}   
  \bibliographystyle{unsrt}

\end{document}